\documentstyle[aps,twocolumn,epsf]{revtex}

\newcommand{\be}{\begin{equation}}
\newcommand{\ee}{\end{equation}}
\newcommand{\bea}{\begin{eqnarray}}
\newcommand{\eea}{\end{eqnarray}}
\def\ep{\epsilon}

\def\bb{C$_{60}$}
\def\abb{$A_3$C$_{60}$}
\def\to{\rightarrow}
\def\dt{\Delta\tau}
\def\ndeg{N_{\rm deg}}

\begin{document}
\draft

\twocolumn[\hsize\textwidth\columnwidth\hsize\csname @twocolumnfalse\endcsname

\title{Multi-Orbital Hubbard Model in Infinite Dimensions: 
Quantum Monte Carlo Calculation
}
\author{
J. E. Han$^a$,
M. Jarrell$^b$,
D. L. Cox$^c$
}
\address{
$^a$-Max-Planck Institut f\"ur Festk\"operforschung, Heisenbergstr.1,
D-70569, Stuttgart, Germany \\
$^b$-Department of Physics, University of
Cincinnati, Cincinnati, Ohio  45221\\
$^c$-Department of Physics, University of
California, Davis, California 95616 
}\maketitle

\widetext
\begin{abstract}
\noindent
Using Quantum Monte Carlo we compute thermodynamics and spectra for 
the orbitally degenerate Hubbard model in infinite spatial dimensions. 
With increasing 
orbital degeneracy we find in the one-particle spectra: broader 
Hubbard bands (consistent with increased kinetic energy), a 
narrowing Mott gap, and increasing quasi-particle spectral weight.  
In opposition, Hund's rule exchange coupling decreases the
critical on-site Coulomb energy for the Mott transition. 
The metallic regime resistivity for two-fold degeneracy is 
quadratic-in-temperature at low temperatures.
\end{abstract}

]

\narrowtext

The Hubbard Hamiltonian, the simplest model for strongly interacting 
many-electron systems, has been extremely popular in that, despite its 
simplicity, the model is considered to capture essential physics in 
electronic systems, ranging from a metal-insulator transition (MIT) and
associated antiferromagnetism  to possible $d$-wave superconductivity. 
Although most of the real systems displaying these phenomena have orbital 
degrees of freedom, most theoretical works have concentrated on the 
orbitally non-degenerate model for simplicity. Recently, with the
advent of the colossal magnetoresistance materials~\cite{millis},
the proposed triplet pairing superconductivity in 
Sr$_2$RuO$_4$~\cite{rice:1},  and alkali-doped
fullerides~\cite{gelfand,gunnar:c60}, 
attention has been turned to the multi-orbital 
Hubbard model (MOHM)~\cite{koch,kajueter,rozenberg}.

With on-site orbital degeneracy 
$\ndeg>1$, one has to consider two additional aspects of the
problem: first, the effect of orbital degrees of freedom on
the mobility of electrons and second, the role of 
on-site exchange interactions between degenerate orbitals. 
In this paper, we 
concentrate on the MIT in MOHM motivated particularly by alkali-doped
fullerides, \abb\ ($A$=K, Rb, Cs etc.).  We have studied
the problem using Quantum Monte Carlo in a $d=\infty$ model on both
Bethe and 
hypercubic lattices, with a focus on dynamical properties (obtained by
analytic continuation of imaginary time data using the Maximum Entropy
method [MEM]).  We present the 
first systematic calculation of one-electron 
spectral functions at and away from
particle-hole symmetry and the first calculation of the optical conductivity
and d.c. resistivity for the two-orbital model.  

We find that the upper
and lower Hubbard bands display a considerable broadening for 
$\ndeg=2,3$ relative to $\ndeg=1$, which supports the
idea of Gunnarsson, Koch, and Martin\cite{koch} and Lu\cite{lu} that
orbital degeneracy serves to increase the effective hopping matrix
element and thereby mitigate the efficacy of correlations in driving a
MIT\cite{lof}.  
We also find that the quasiparticle spectral weight increases with 
$\ndeg$. We find that the inclusion of Hund's rule exchange serves to decrease
the effective hopping, thereby decreasing again the critical value of
on-site Coulomb interaction $U=U_c$ needed for the MIT, according to the
approximate expression
\be
  U_c(\ndeg,J)\approx \sqrt{\ndeg}U_c(1,0)-\ndeg J,
\label{eq:uc}
\ee
where $J$ is the Hund's rule exchange energy.
Finally, we present a
calculation of the optical conductivity $\sigma(\omega,T)$ 
and resistivity $\rho(T)$ for the $\ndeg=2$
case which gives a $T^2$ behaviour in $\rho(T)$ for $T\to 0$.

We were motivated to carry out this study by consideration of \abb.
Even though the local density approximation 
predicted well the band-width and the band-gap of this 
system, it has been suggested that the strong Coulomb 
repulsion on the C$_{60}$ site could drive the system to a strongly 
correlated metal or even to a Mott-insulator~\cite{lof}.
Since the band-width of \abb\ (0.4 eV) is more than two times smaller
than the on-site Coulomb repulsion (1-1.5 eV), one can naively
predict that the system should be a Mott-insulator, if interpreted in 
terms of an orbitally nondegenerate Hubbard model. 
However, the alkali-doped fullerides are metals despite the
large resistivity at zero temperature~\cite{xiang}.

It has been suggested that the metallic \abb\ can be 
understood by introducing orbital degeneracy in the Hubbard 
model~\cite{koch,lu}. The MIT in the MOHM has been studied by
the Gutzwiller approximation~\cite{lu}, where Lu showed that
the critical value for the on-site Coulomb repulsion strength 
$U_c(\ndeg)$ for MIT at half-filling increases as
$(\ndeg+1)U_c$ with the orbital degeneracy $\ndeg$.
This explains why alkali-doped
fullerides ($\ndeg=3$) can be metallic despite the large 
intra-molecular Coulomb
repulsion and small conduction band-width~\cite{gelfand}.
Later, Gunnarsson
{\em et al.}~\cite{koch} argued that $U_c(\ndeg)$ behaves
as $\sqrt{\ndeg}$, 
in a much weaker fashion than the linear dependency
of Ref.~\cite{lu}. The electron hopping is argued to be effectively 
enhanced with the orbital degeneracy due to the anti-ferromagnetic
correlation between nearest sites, which is also supported by their
numerical calculation. 

To carry out our studies, 
we have used the dynamical mean field theory (DMFT)~\cite{dmft}, where
a lattice Hamiltonian is rigorously mapped to an effective impurity
model. This approximation, which is exact for infinite spatial dimension
or coordination number, is reasonable for modelling the
\bb-derived $t_{1u}$ band due to the large coordination number, 12,
between the \bb\ molecules located at face-centered-cubic sites. 
Furthermore, with the on-site Coulomb interaction, the impurity
Hamiltonian from the DMFT is expected to retain the essential physics 
of the MIT. The multi-orbital Hubbard Hamiltonian reads
\be
  H = -\!\!\!\sum_{ij,m\sigma}\!\!\left(t_{ij}c^\dagger_{im\sigma}c_{jm\sigma}
	+ h.c.\right) + \ep_d\sum_{im\sigma}n_{im\sigma} + H_C,
\ee
where $t_{ij}$ is the hopping matrix element between different sites 
$i$ and $j$, $m$ the orbital index, $\sigma$ the spin index,
$\ep_d$ the $d$-level energy as the center of the non-interacting band,
$H_C$ the interaction Hamiltonian for Coulomb energy. 
For simplicity, we have
neglected inter-band hopping since we concentrate on the local
properties.  

The Coulomb interaction term is written~\cite{dworin} as
\bea
  H_C & = & 
	(U+J)\sum_m n_{im\uparrow}n_{im\downarrow}+
	(U-J)\!\!\sum_{im<m'\sigma}\!\!
	n_{im\sigma}n_{im'\sigma} \nonumber \\ 
	& & +
	U\!\!\sum_{im\neq m'}\!\!n_{im\uparrow}n_{im'\downarrow} +
	J\!\!\sum_{m<m'}\!c^\dagger_{im'\uparrow}
	c^\dagger_{im\downarrow}c_{im'\downarrow}c_{im\uparrow}.
	\label{eq:hint}
\eea
Here, we discard the last term proportional to $J$ for
computational convenience although, for full rotational symmetry, all the
terms in the above Hamiltonian are needed. 
The truncated Hamiltonian still captures the Hund's rule physics,
and we believe that the form of Eq. (\ref{eq:uc})
remains valid, possibly with a
small correction to the term proportional to $J$.

The effective impurity model in the DMFT is studied by the Hirsch-Fye
algorithm~\cite{fyehirsch} where the partition function is reformulated 
using a discrete path integral with an
imaginary-time interval which we set to $\dt=1/3$. For the free
density of states (DOS) in the DMFT, a semi-circular DOS (
corresponding to the infinite coordination number Bethe lattice),
$D(\ep)={2\over \pi}\sqrt{1-\ep^2}$ with band-width $W$=2, is used for 
calculations of thermodynamic quantities such as occupation numbers.
A Gaussian DOS (corresponding to the infinite dimensional hypercubic 
lattice), $D(\ep)={1\over \sqrt{\pi}}\exp(-\ep^2)$ with $W$=$\sqrt{2}$, 
is used for one-particle spectral functions.
We normalize $t_{ij}=t$ by $t=t^*/2\sqrt{d}$, as the dimensionality $d$
goes to infinity.  
All energies here are measured in units
of $t^*$ for the Gaussian DOS, and in units of the bandwidth for
the semi-circular DOS.

\begin{figure}
\centerline{\epsfxsize=3.375in \epsffile{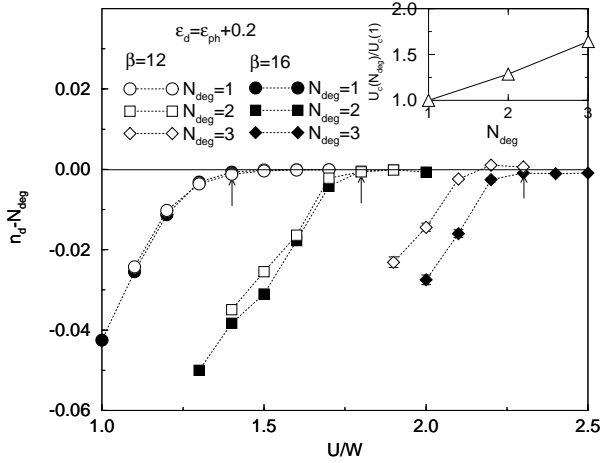}}
\caption[]{
The occupancy $n_d$ versus the Coulomb repulsion $U$ at fixed
$\ep_d=\ep_{ph}-(\ndeg-{1\over 2})U+0.2$. The critical value of $U$
(marked with arrows at $T=1/16$) at
the cross-over between insulating and metallic solutions increased with
$\ndeg$. The temperature, $T=1/16$, is not low enough for $\ndeg=3$
case to equate the $U_c=2.3W$ as the zero-temperature quantum critical
point. Due to the weak functional dependency of $\sqrt{\ndeg}$ and
small
$\ndeg$, it is not clear that $U_c(\ndeg)=\sqrt{\ndeg}U_c(1)$.
(See inset.) Lines are guide to the eye.
}
\label{fig:mottNxUx}
\end{figure}

At $J=0$, we identified the MIT for $\ndeg=1,2,3$ near
half-filling by computing the occupation numbers at $\ep_d$ 
away from the particle-hole symmetric value, 
{\em i.e.,} $\ep_d=\ep_{ph}+0.2$ where $\ep_{ph}=(N_{\rm
deg}-{1\over 2})U+(1-{N_{\rm deg}\over 2})J$.
FIG.~\ref{fig:mottNxUx}
shows the occupation numbers versus the Coulomb repulsion with
the semi-circular DOS. The occupation numbers and the chemical 
potentials are plotted with respect to the particle-hole
symmetric values. 
The temperatures are at $T=1/12,1/16$.
When the system is insulating with the chemical
potential inside an insulating gap, the occupation number does not change
with a slight shift of chemical potential from $\ep_{ph}$. 
As will be illustrated later, the Mott insulating-gap 
manifests as a plateau in $n_d$ vs $\epsilon_d$.
The critical Coulomb repulsion, $U_c$, is read off from the 
point where the
occupation number $n_d$ at $\ep_d=\ep_{ph}+0.2$
deviates from $\ndeg$ in FIG.~\ref{fig:mottNxUx}.

$U_c(\ndeg)$ increases with the orbital degeneracy, in agreement with
Ref.~\cite{koch}, although the data are not sufficiently precise
to conclude that the
insulating gap is proportional to $\sqrt{\ndeg}$ from small $\ndeg$
and $U/W$. For $\ndeg=3$, the occupation numbers have not fully
converged yet for $T=1/16$ and we obtain $U_c/W\geq 2.3$.
Compared with \abb, this value of $U_c$ puts the experimental ratio near the edge
of MIT transition. Therefore, it is theoretically possible, if not
decisively proven, that \abb\ can be described as a strongly correlated
metal despite being at half filling.

\begin{figure}
\centerline{\epsfxsize=3.375in \epsffile{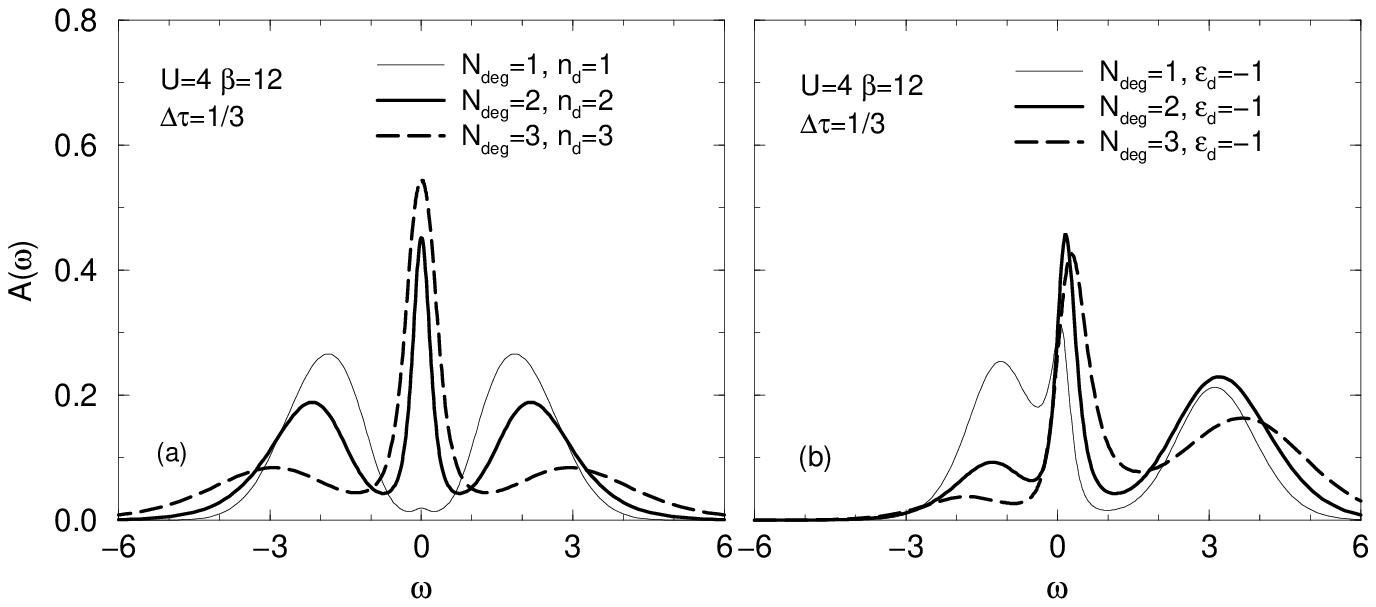}}
\caption[]{
Spectral functions for $\ndeg=1,2,3$ at $U=4$ and $J=0$.
(a) The effective band-width of the charge excitation peaks (nearly
at $\pm U/2$ at half-filling) is broadened with increasing
$\ndeg$ due to the enhanced hopping for one-particle (hole) doped states. 
The quasi-particle (QP) 
peak becomes dominant for large $\ndeg$, as more weight of
charge excitation peaks is transferred to the QP  peak.
(b) At a fixed $\ep_d=-1$, all three cases have 
total occupations near 0.9 and the integrated weight up to the chemical
potential is roughly $1/\ndeg$.
}
\label{fig:dosNx}
\end{figure}

The QMC Green's function along imaginary time is analytically continued
for the one-particle spectral function by MEM~\cite{mem}.
One-particle spectral functions at and away from half-filling
are shown in FIG.~\ref{fig:dosNx} when $T=1/12$ and $J=0$. 
In plot (a), at half-filling, the upper
(UHB) and lower Hubbard bands (LHB) are split by the Coulomb
interaction strength $U$. Note that the width of these peaks 
increases with the orbital degeneracy, in the same
fashion that the electron hopping is enhanced for conduction of the 
charge excitation. For the non-degenerate model ($\ndeg=1$), the
spectral weight at zero frequency is very small, hence a sign of
an insulating pseudo-gap (the system cannot become a true insulator
due to the non-vanishing tail of the Gaussian free DOS). With increasing 
$\ndeg$, the MOHM evolves to solutions with more metallic
character.

The peak positions of the UHB and LHB are progressively shifted further
away from $\pm U/2$ as $\ndeg$ increases. These shifts, roughly
proportional to $\ndeg t^2/U$ for $U$ sufficiently large, can be
understood in terms of the perturbational contributions to energy levels
with occupation numbers $\ndeg-1$, $\ndeg$ and $\ndeg+1$, respectively. 
When $t \approx U$, 
one should also take into account this shift in
addition to the band broadening at the MIT, which can complicate the 
argument of Gunnarsson, Koch, and Martin\cite{koch} that led to a 
$\sqrt{\ndeg}$ dependence to $U_c$.
 
The central peaks for $\ndeg=2,3$ are the quasi-particle (QP)
excitation which emerges from the scattering of the spin degrees of
freedom of conduction electrons. The width of the QP peak defines the
new low temperature energy scale (renormalized Fermi energy).
Since the spectral weight from the UHB and LHB is
expected to be larger at $\omega=0$ for larger $\ndeg$ due to the
broadening mentioned in a previous paragraph, 
the QP peak accordingly has larger weight at the 
chemical potential. This trend goes with enhanced
metalicity for increasing $\ndeg$. We will discuss more
the dependency of the QP weight upon orbital degeneracy in 
relation to the electronic resistivity.
 
FIG.~\ref{fig:dosNx} (b) shows the one-particle spectral functions away
from half-filling at $U=4,\ep_d=-1$ for all $\ndeg$. 
In the
paramagnetic phase, the total occupation number is evenly distributed
among the orbitals and the spectral weight of the LHB roughly goes as
$1/\ndeg$.
Since $|\ep_d|\ll U$ and
the doubly occupied configuration exists in small quantum weight, 
total occupation number is not very sensitive to the
orbital degeneracy.

\begin{figure}
\centerline{\epsfxsize=3.375in \epsffile{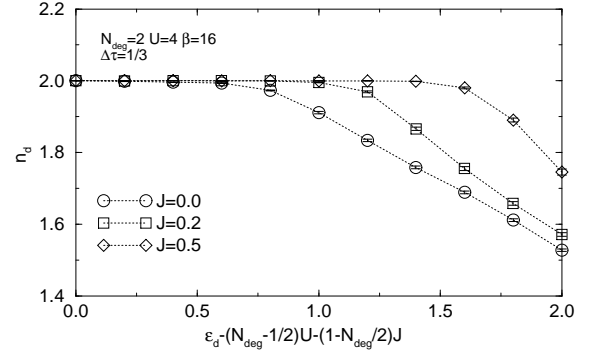}}
\caption[]{
$n_d$ vs. $\ep_d-\ep_{ph}$ at $J=0.0,0.2,0.5$. For positive $J$
the onset of a metallic solution is shifted to larger values of $\ep_d$
from the particle-hole symmetric parameter, $\ep_{ph}$. The shift is
roughly linear in $J$, leading to a suggestion that the on-site Coulomb
interaction is responsible with $\Delta U(J)=\ndeg J$.
}
\label{fig:mottN2Jx}
\end{figure}

The behavior of the MIT with a finite Hund's rule coupling $J$ is shown
in FIG.~\ref{fig:mottN2Jx}. 
When the exchange interaction $J$ in Eq. (\ref{eq:hint}) is included,
the Coulomb repulsion is effectively increased, favoring the insulating
phase. Numerical calculation confirms that the insulating phase is
expanded with finite positive $J$. Figure~\ref{fig:mottN2Jx} shows the
plot of $n_d$ vs. $\ep_d-\ep_{ph}$ with $J=0.0,\;0.2,\;0.5,\;1.0$ at
$\ndeg=2\quad\mbox{and}\quad\beta=16$. The total occupation number
$n_d$ begins to deviate from the half-filling value $\ndeg$ when the
system become metallic. $\ep_d$'s where the metallic
phases begin are pushed to higher values with $\Delta\ep_d$
approximately linear in $J$. In other words, the insulating gap has
increased with finite $J$.

The phenomena can be easily understood by considering the change in
the on-site energy at half-filling. 
The effective on-site Coulomb repulsion $U_{eff}$ can be computed
as $E(\ndeg+1)+E(\ndeg-1)-2E(\ndeg)$ from the on-site Hamiltonian Eq.
(\ref{eq:hint}), which leads to
\be
  U_{eff}=U(\ndeg,J=0)+\ndeg J,\quad
 \mbox{linear\ in\ }\ndeg.
\label{eq:nj}
\ee
Also, we have to remember that the hopping of electrons leaves behind
a trail of broken Hund's rule multiplets. Therefore, there would be
additional effective repulsion due to hopping, which is ignored
in Eq. (\ref{eq:nj}). For \abb, $J$ is
one order of magnitude smaller than $U$~\cite{lof}, the Hund's rule
coupling plays less important role than $\sqrt{\ndeg}U_c(1,0)$ in Eq.
(\ref{eq:uc}). For systems with larger $J$, one should take into
account the last term in Eq. (\ref{eq:hint}) to properly describe the
magnetic correlation.

\begin{figure}
\centerline{\epsfxsize=3.375in \epsffile{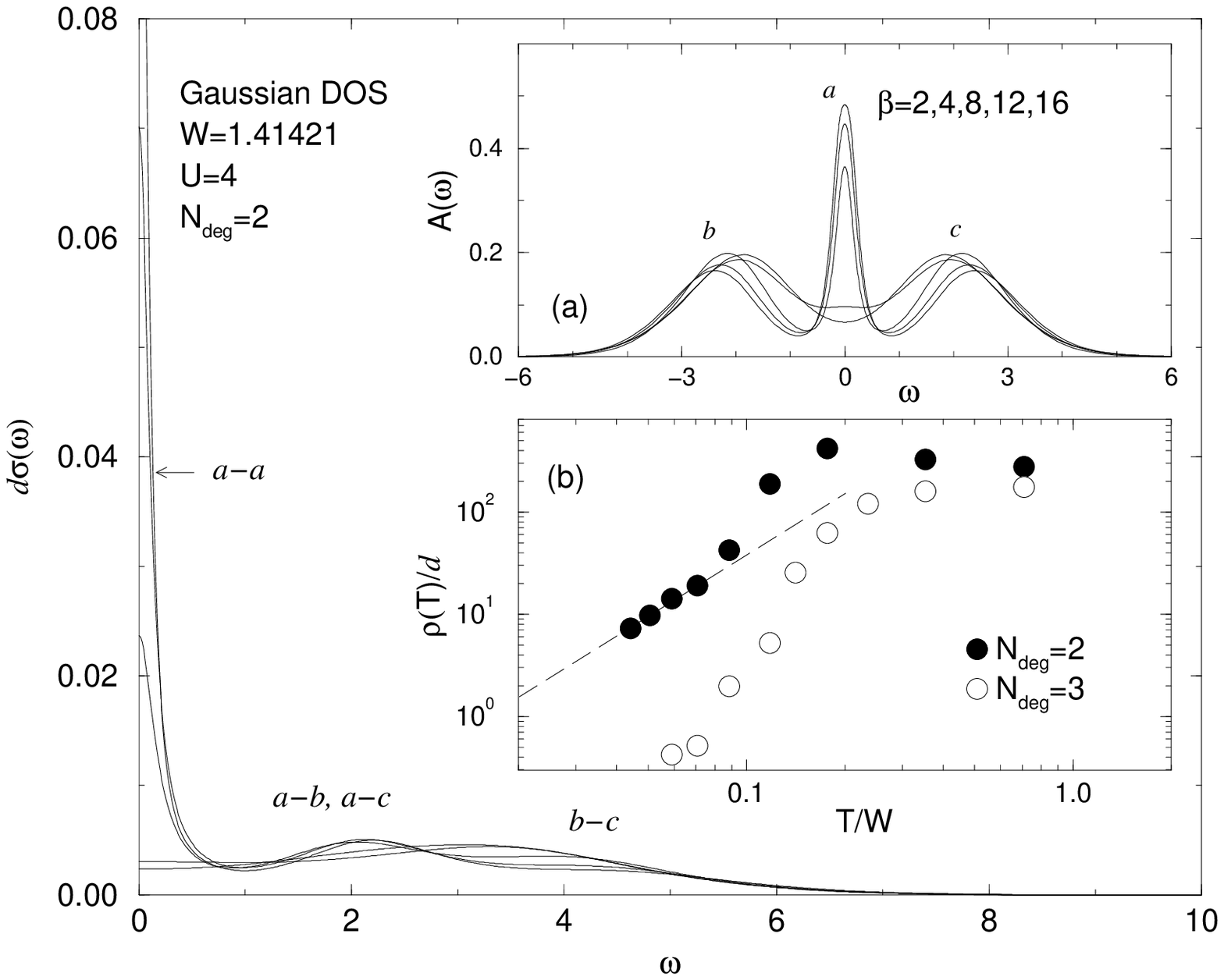}}
\caption[]{
Optical conductivity and resistivity for $\ndeg=2$ at $U=4$,
$\ep_d=-1$. The optical conductivity is made up of three excitations
arising from transitions
between three structures in the spectral functions
plotted in the inset (a). The spectral functions at
$\beta=2,4,8,12,16$ have QP peak, LHB, UHB marked as $a$, $b$,
$c$, respectively. The Drude peak, arising from the scattering within
the QP peak, becomes stronger as $T$ lowers. 
The d.c. resistivity $\rho(T)$ for the doubly degenerate-band
Hubbard model 
(filled circle) illustrates that 
the Fermi liquid behavior (dashed line) holds 
at half-filling below the critical $U$ value. The triply degenerate
case (empty circle) has much lower values for the resistivity.
The unit of $\rho(T)/d$ in the figure is $\mu\Omega{\rm cm}$~\cite{unit}.
}
\label{fig:dsigma}
\end{figure}

Finally, we discuss the transport properties of the MOHM. 
From the analytically continued one-particle spectral function, we can
extract the electron scattering rate (imaginary part of the self-energy) 
and thereby 
calculate the optical conductivity, $\sigma(\omega)$,
within the DMFT~\cite{pruschke}, as shown in FIG.~\ref{fig:dsigma}. 
The inset (a) shows the evolution of
spectral functions for a half-filled doubly degenerate model
at inverse temperatures $\beta=2,4,8,12,16$, where the curve 
with the large QP peak (labelled as $a$ in the figure) corresponds 
to low $T$.
The optical conductivities shown in the main plot are from the 
same set of temperatures. The optical conductivity
is computed by convoluting the one-particle spectral functions and
is therefore characterized by transitions between various 
single-particle excitations, as indicated in the figure.
The Drude peak (marked as $a-a$), resulting from the
scattering {\em within} the QP peak $a$, diverges as the temperature
goes to zero. Note that, since the leading order of $\sigma(\omega)$
is $O(1/d)$, we have plotted $d\sigma(\omega)$.

The resistivity $\rho(T)$, plotted in inset (b),
is calculated from the $\omega\to 0$ limit of the optical conductivity,
$\rho(T)^{-1}=\lim_{\omega\to 0}\sigma(\omega)$. 
The unit of $\rho(T)/d$ is $\mu\Omega{\rm cm}$~\cite{unit}.
The data shown here are for half-filled Hubbard models
with double (filled circle) and triple (open circle) 
orbital degeneracy at $U=4$ using the Gaussian free DOS with 
effective band width $W=\sqrt{2}$.
The resistivity for doubly degenerate model shows a cross-over
to a Fermi liquid behavior, $\rho(T)\propto T^2$, at low $T$ from 
a resistivity of insulating character at high $T$. For the triply
degenerate model, we could not determine the low temperature
law of resistivity due to noise in QMC data.

It should be noted that $\rho(T)$ for $\ndeg=3$ is about one 
order of magnitude smaller than $\rho(T)$ for $\ndeg=2$, 
in contrast to
the weak $\ndeg$-dependency in the low-$U$ perturbation 
theory~\cite{gunnar:priv}. Although not discussed in this paper, 
the magnitude of $\rho(T)$ seems to strongly depend upon the shape
of band edge of free DOS. For example, the resistivity for 
semi-circular DOS is order(s) of magnitude larger than for Gaussian
DOS at the same $U/W$ ratio. A systematic investigation
of resistivity on the orbital degeneracy and DOS and its 
comparison with experiments are left for further studies. 

We acknowledge an initial inspiration from T.M Rice to study this
subject, along with useful discussions with H. R. Krishna-murthy, O.
Gunnarsson.
This work was supported in part by the U.S. Department of Energy, 
Office of Basic Energy Sciences, Division of Materials Research
(J.H. and D.L.C.) and by National Science Foundation grant
DMR-9704021 and DMR-9357199 (M.J.).  
Supercomputer time was provided by the Ohio Supercomputer Center.

\begin{references}

\bibitem{millis}
A.~J. Millis, R.~Mueller, and B.~I. Shraiman,
\newblock Phys. Rev.~B {\bf 54}, 5389 (1996).

\bibitem{rice:1}
T.~M. Rice and M.~Sigrist,
\newblock J. Phys.: Condens. Matter {\bf 7}, L643 (1995),
\newblock References therein.

\bibitem{gelfand}
M.~P. Gelfand,
\newblock ``Alkali fullerides: Theoretical perspectives, progress and
  problems'',
\newblock in {\em Superconductivity Review} Vol.~1, p. 103, Gordon and Breach
  Science Publishers, 1994.

\bibitem{gunnar:c60}
O.~Gunnarsson,
\newblock Rev. Mod. Phys. {\bf 69}, 575 (1996); private communication.

\bibitem{koch}
O.~Gunnarsson, E.~Koch, and R.~M. Martin,
\newblock Phys. Rev.~B {\bf 54}, R11026 (1996).

\bibitem{lu}
J.~P. Lu,
\newblock Phys. Rev.~B {\bf 49}, 5687 (1994).

\bibitem{lof}
R.~W. Lof, M.~A. van Veenendaal, B.~Koopmans, H.~T. Jonkman, and G.~A.
  Sawatzky,
\newblock Phys. Rev. Lett. {\bf 68}, 3924 (1992).

\bibitem{kajueter}
H.~Kajueter and G.~Kotliar,
\newblock Int. J. Mod. Phys. B {\bf 11}, 729 (1997).

\bibitem{rozenberg}
M.~J. Rozenberg,
\newblock Phys. Rev.~B {\bf 55}, R4855 (1997).


\bibitem{xiang}
X.-D. Xiang, J.~G. Hou, V.~H. Crespi, A.~Zettl, and M.~L. Cohen,
\newblock Nature {\bf 361}, 54 (1993).


\bibitem{dmft}
W.\ Metzner and D.\ Vollhardt, Phys.\ Rev.\ Lett.\ {\bf 62}, 324 (1989);
Th.\ Pruschke, M.\ Jarrell and J.K.\ Freericks, Adv.\ in Phys.\ {\bf{42}}, 
187 (1995);
A.~Georges, G.~Kotliar, W.~Krauth, and M.~J. Rozenberg,
\newblock Rev. Mod. Phys. {\bf 68}, 13 (1996).

\bibitem{dworin}
L.~Dworin and A.~Narath,
\newblock Phys. Rev. Lett. {\bf 25}, 1287 (1970).


\bibitem{fyehirsch}
R.~M. Fye and J.~E. Hirsch,
\newblock Phys. Rev.~B {\bf 38}, 433 (1988).

\bibitem{mem}
M.\ Jarrell and J.E.\ Gubernatis, Physics Reports 
Vol.\ {\bf{269}} \ No.3, p133-195, (May, 1996).

\bibitem{pruschke}
T.~Pruschke, D.~L. Cox, and M.~Jarrell,
\newblock Phys. Rev.~B {\bf 47}, 3553 (1993).

\bibitem{unit} 
$\rho(T)/d=
(\pi e^2/ a^{d-2} \hbar)f(T/t^*,U/t^*,\epsilon_d/t^*)$ with the 
lattice constant $a$, and dimensionless $f(T/t^*,U/t^*,\epsilon_d/t^*)$. 
Here $a$ is set to $1 \AA$ and $d$ to 3.

\bibitem{gunnar:priv}
O.~Gunnarsson,
\newblock Private Communication.

\end{references}
\end{document}